\documentclass[preprint,review,12pt,sort&compress]{elsarticle}
\usepackage[margin=1.18 in]{geometry}
\usepackage{multirow}
\usepackage[usenames, dvipsnames]{color}
\definecolor{UW}{RGB}{64, 38, 96}

\usepackage{amssymb}

\usepackage{float}

\journal{Polymer}

\begin{document}

\begin{titlepage}

\clearpage\thispagestyle{empty}



\noindent

\hrulefill

\begin{figure}[h!]

\centering

\includegraphics[width=1.5 in]{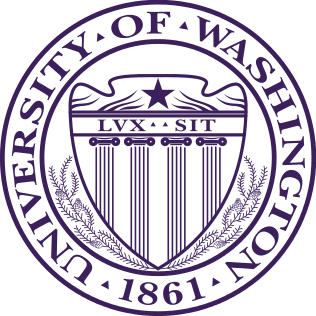}

\end{figure}

\begin{center}

{\color{UW}{

{\bf A\&A Program in Structures} \\ [0.1in]

William E. Boeing Department of Aeronautics and Astronautics \\ [0.1in]

University of Washington \\ [0.1in]

Seattle, Washington 98195, USA

}

}

\end{center} 

\hrulefill \\ \vskip 2mm

\vskip 0.5in

\begin{center}

{\large {\bf Strength and Cohesive Behavior of Thermoset Polymers at the Microscale: A Size-Effect Study}}\\[0.5in]

{\large {\sc Yao Qiao, Marco Salviato}}\\[0.75in]

{\sf \bf INTERNAL REPORT No. 18-12/03E}\\[0.75in]

\end{center}

\noindent {\footnotesize {{\em Submitted to Polymer \hfill December 2018} }}

\end{titlepage}

\newpage

\begin{frontmatter}


\cortext[cor1]{Corresponding Author, \ead{salviato@aa.washington.edu}}

\title{Strength and Cohesive Behavior of Thermoset Polymers at the Microscale: A Size-Effect Study}



\author[address]{Yao Qiao}
\author[address]{Marco Salviato\corref{cor1}}

\address[address]{William E. Boeing Department of Aeronautics and Astronautics, University of Washington, Seattle, Washington 98195-2400, USA}

\begin{abstract}
\linespread{1}\selectfont
This study investigated, experimentally and numerically, the fracturing behavior of thermoset polymer structures featuring cracks and sharp u-notches. It is shown that, even for cases in which the sharpness of the notch would suggest otherwise, the failure behavior of cracked and pre-notched specimens is substantially different, the failure loads of the former configuration being about three times lower than the latter one. To capture this interesting behavior a two-scale cohesive model is proposed. The model is in excellent agreement with the experimental data and its predictions allow to conclude that (a) residual plastic stresses cannot explain the very high failure loads of notched structures; (b) the strength of the polymer at the microscale can be from six to ten times larger than the values measured from conventional tests whereas the fracture energy at the microscale can be about forty times lower; (c) the pre-notched specimens investigated in this work failed when the stress at the tip reached the microscale strength whereas the cracked specimens failed when the energy release rate reached the total fracture energy of the material. The foregoing considerations are of utmost importance for the design of microelectronic devices or polymer matrix composites for which the main damage mechanisms are governed by the strength and cohesive behavior at the microscale.




\end{abstract}


\end{frontmatter}



\section{Introduction}
Thermoset polymers find extensive application across the main engineering fields from e.g. automotive, aerospace and civil engineering to microelectronics \cite{automotive,aerospace,civil,microelectronic}. Thermosets are also the material of choice for the manufacturing of advanced fiber composites \cite{stenzen,pascault} although the demand for recyclability, high manufacturability, and damage tolerance is gradually shifting the focus to thermoplastics \cite{takahashi,blok,shan,plasticsbook}.

Considering the several structural applications of thermosets, understanding the fracturing behavior of these materials is quintessential and has been the subject of extensive research in the past four decades \cite{Knauss,kinloch,Narisawa,argon1,argon,williams,Kwon}. However, while significant progress has been made in the characterization and modeling of crack initiation and propagation, far less attention has been devoted to the damaging and fracturing behavior of thermosets in the presence of sharp notches \cite{xiao,cayard,alicia,souza,kuppu,zappalorto,zappalorto2}. This is an important issue considering that most of the thermoset structures inevitably feature sharp notches. Examples include microelectronic devices in which sharp geometrical features are ubiquitous or fiber composites where the matrix is subjected to the stress concentration induced by the fibers.

Narisawa \emph{et al.} \cite{Narisawa} investigated the fracturing surface morphology of epoxy resin and showed that an internal crack may nucleate at the boundary between the plastic and elastic regions, thus generating a Fracture Process Zone that affects the fracturing behavior. A comprehensive analysis on the failure mechanism of epoxy resin was provided by Kinloch \emph{et.al} \cite{kinloch}, who proposed a quantitative model accounting for the blunting at the notch tip prior to the onset of crack initiation. Several potential toughening mechanisms for polymer structures weakened by cracks and sharp notches were discussed by Argon \emph{et.al} \cite{argon1,argon} whereas the effect of different pre-notching methods on the fracturing behavior of polymers was studied in \cite{xiao,cayard,alicia,souza}. In these contributions it was proposed that residual plastic stresses induced by the manufacturing of the notch may lead to very high values of apparent fracture toughness of the polymer.


According to the foregoing contributions, the failure behavior of cracked and pre-notched specimens is substantially different, even when the sharpness of the notch and Linear Elastic Fracture Mechanics (LEFM) would suggest otherwise. The present work attempts to clarify this difference leveraging computational cohesive fracture mechanics and size effect testing of both pre-cracked and pre-notched Single Edge Notch Bending (SENB) specimens. A \textit{two-scale} cohesive law is proposed and shown to provide an excellent agreement with the experimental data. Leveraging the model, not only the notch mechanics of thermosets is clarified but also unprecedented insight on the strength and cohesive behavior of the polymer at the microscale is obtained. It is shown that the microscale strength can be from six to ten times larger than the values estimated from macroscale tests. In contrast, the fracture energy at the microscale is estimated to be roughly forty times lower than the values obtained from traditional fracture tests.

The results of this work disprove the hypothesis, largely accepted in the literature \cite{xiao,cayard,alicia,souza,kuppu}, that residual plastic stresses generated by the notching process are the cause of the different behavior of notched and cracked structures. More importantly, the two-scale cohesive model proposed in this study represents a first step towards the better understanding of the cohesive behavior of thermosets at the microscale. This information is quintessential for the formulation of accurate computational models for microelectronic devices or the damaging and fracturing behavior of the matrix in fiber composites. Further, the novel insight on the cohesive behavior can pave the way for the development of new nanomodification strategies targeting specifically the enhancement of the behavior at the microscale \cite{Zappalorto1,salviato5,Zappalorto3,salviato2,quaresimin}.

\section{Materials and methods} 
\subsection{Material preparation}
\label{sec:materialsandpreparation}
Following \cite{Cory,Yao0}, the thermoset polymer used in this work was composed of an $\textnormal{EPIKOTE}\textnormal{\texttrademark}$ Resin  $\textnormal{MGS}\textnormal{\texttrademark}$ and an $\textnormal{EPIKOTE}\textnormal{\texttrademark}$ curing Agent $\textnormal{MGS}\textnormal{\texttrademark}$ RIMH 134-RIMH 137 (Hexion \cite{hexion}) combined in a 100:32 ratio by weight.  

The epoxy and hardener were mixed for 10 minutes and degassed for 20 minutes in a vacuum trap using a Vacmobile mobile vacuum system \cite{pump} in order to remove any air bubbles. After degassing,  the mixture was poured into silicone molds made of RTV silicone from TAP Plastics \cite{tapplastics} to create geometrically-scaled specimens with consistent sizes. Finally, the resin was allowed to cure at room temperature for approximately 48 hours and then post-cured in an oven for $4$ hours at $60$ $^{\circ}$C.

\subsection{Specimen preparation}
 Several previous investigations on Compact Tension (CT) and Single Edge Notch Bending (SENB) specimens made of thermoset polymers have shown that the method used to create the notch has a significant influence on the failure behavior and the ultimate load \cite{xiao,cayard,alicia,souza,kuppu}. In particular, it has been shown that specimens with pre-cracks created by tapping may exhibit values of the fracture toughness from $5$ to $20$ times lower than the ones estimated by testing pre-notched specimens, regardless of the way the pre-notch is made (e.g insertion of a teflon sheet or micro-sawing). This discrepancy cannot be explained by Linear Elastic Fracture Mechanics (LEFM) since, considering the very small ratios between the notch tip radius and the notch depth, this theory predicts the same failure load in case of the pre-notch or the pre-crack \cite{lazzarin,lazzarin2,atzori,atzori2}. 
 
\begin{figure} [!ht]
\center
\includegraphics[scale=1.1]{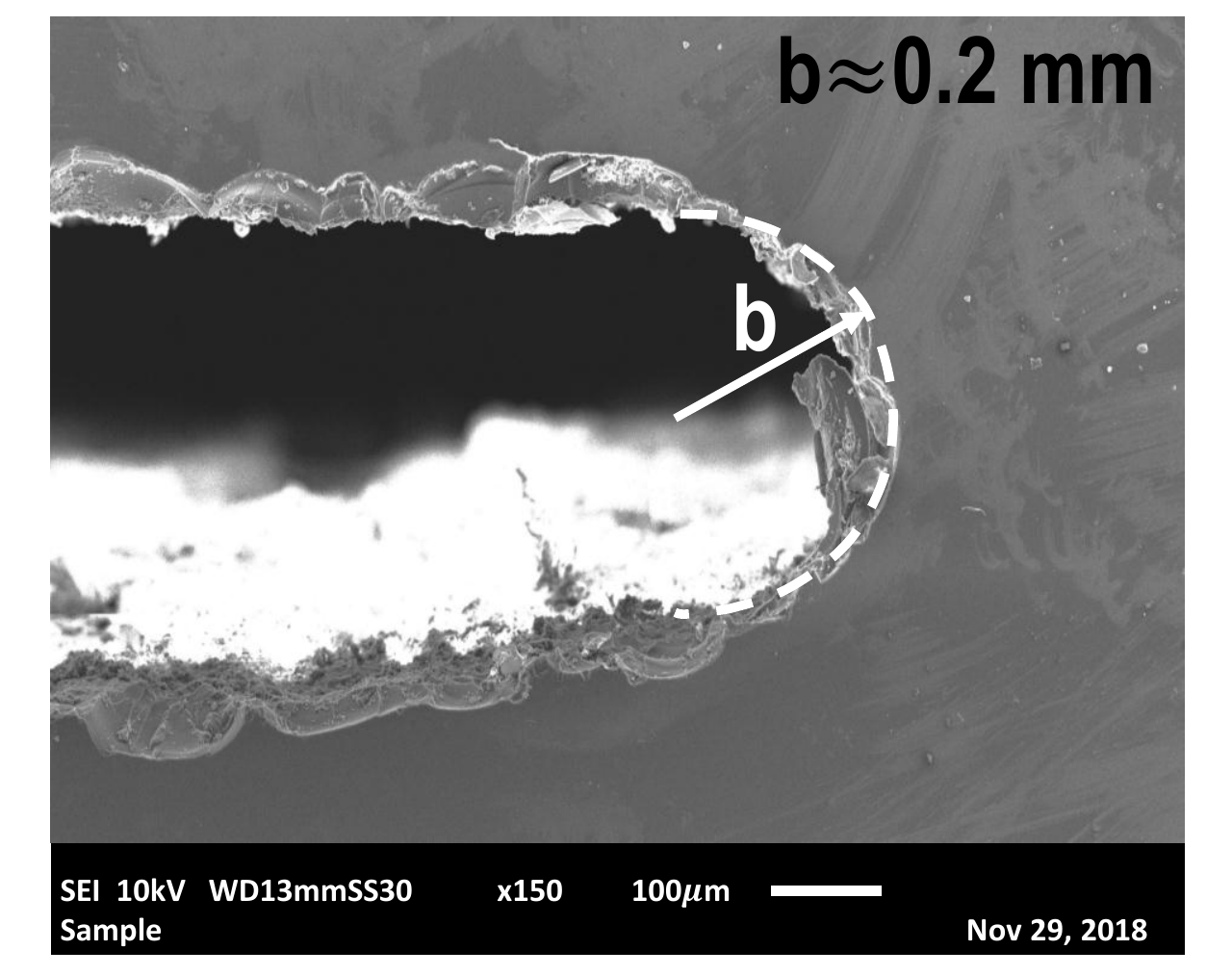}
\caption{Notch tip geometry from Scanning Electron Microscopy (SEM).}
\label{fig:notch}
\end{figure}

 To clarify the foregoing differences and to provide an objective investigation of the fracture properties of the polymer both pre-cracked and pre-notched Single Edge Notch Bending (SENB) specimens were investigated in this work. The pre-notched specimens were created by means of a 
$0.3$ mm wide diamond coated saw leading to a 0.4 mm wide notch as illustrated in Figure \ref{fig:notch} which shows a magnification of the notch tip by means of Scanning Electron Microscopy (SEM). The pre-cracked specimens were created through a two-stage process. The first step consisted in creating a notch about one quarter of the specimen width by means of the $0.3$ mm wide diamond coated saw. Then, tapping leveraging a sharp razor blade followed to create the last portion of the crack. 

\subsubsection{Pre-cracked specimens}
\begin{figure} [!ht]
\center
\includegraphics[scale=0.48]{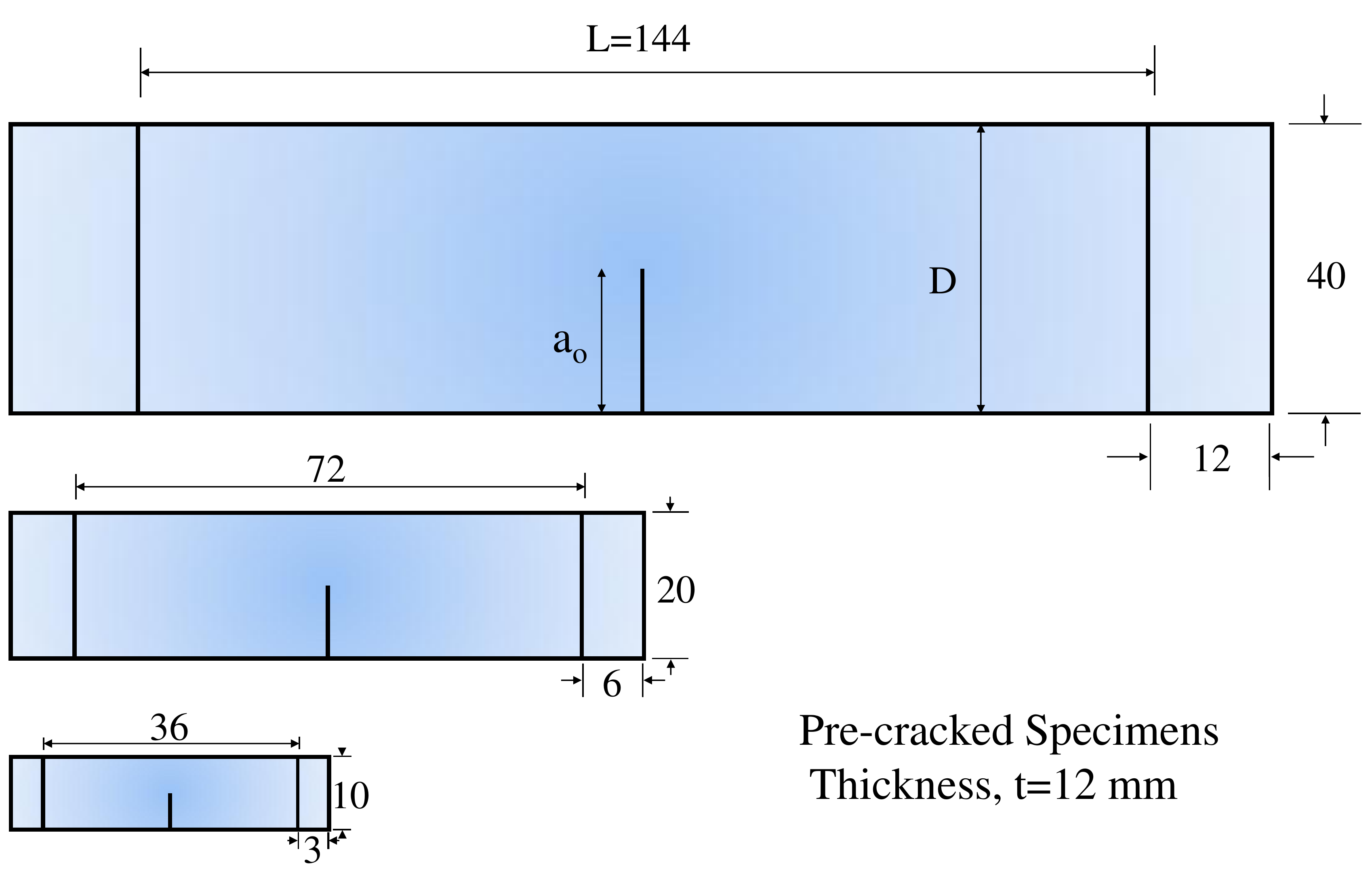}
\caption{Geometry of the pre-cracked specimens. Units: mm.}
\label{fig:precrackgeometry}
\end{figure}

The design of the Single Edge Notch Bending (SENB) specimens was based on ASTM D5045-99 \cite{ASTM_SENB}. In order to study the scaling of the fracturing behavior of pre-cracked specimens, geometrically-scaled specimens of three different sizes were prepared as illustrated in Figure \ref{fig:precrackgeometry}. The dimensions, scaled as $1:2:4$, were $10\times36$ mm, $20\times72$ mm, and $40\times144$ mm, respectively. The various crack lengths of the specimens were approximately in the range $0.35$$D$ to $0.55$$D$, where $D$ is the width of the specimen. This aspect was very important to guarantee a proper geometrical scaling. The scaling did not involve the thickness, $t$, which was kept about $12$ mm for all the investigated sizes.

\subsubsection{Pre-notched specimens}
\begin{figure} [!ht]
\center
\includegraphics[scale=0.25]{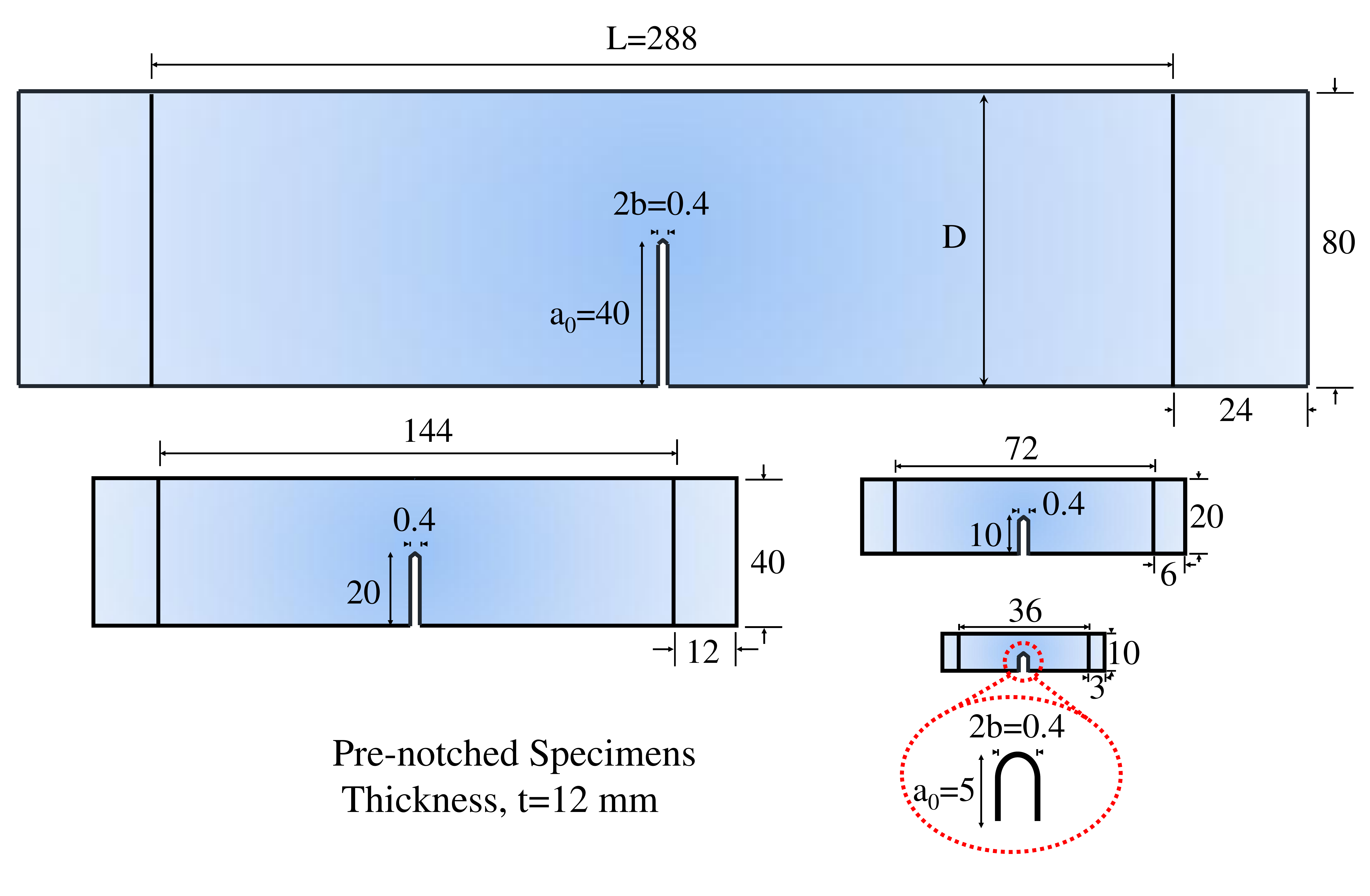}
\caption{Geometry of the pre-notched specimens. Units: mm.}
\label{fig:prenotchgeometry}
\end{figure}

As illustrated in Figure \ref{fig:prenotchgeometry}, geometrically scaled specimens of four different sizes were prepared. The dimensions, scaled as $1:2:4:8$, were $10\times36$ mm, $20\times 72$ mm, $40\times 144$ mm and $80 \times 288$ mm, respectively while the width of the notch was kept the same for all the investigated sizes. Accordingly, the ratio between the depth and the radius of the notch, $a_{0}/b$, was $25$, $50$, $100$, and $200$ respectively. The notch length was always half of the width of the specimen. The thickness, $t$, was kept about $12$ mm for all the investigated sizes.

\subsection{Fracture testing}

Three-point bending tests were performed using a closed-loop, electro-actuated 5585H Instron machine. To avoid viscoelastic effects, the load rate for the three-point bending tests was adjusted for the different sizes to achieve roughly the same average strain rate of $0.2$ \%/min.

\section{Experimental Results}

\begin{figure} [!ht]
\center
\includegraphics[scale=0.7]{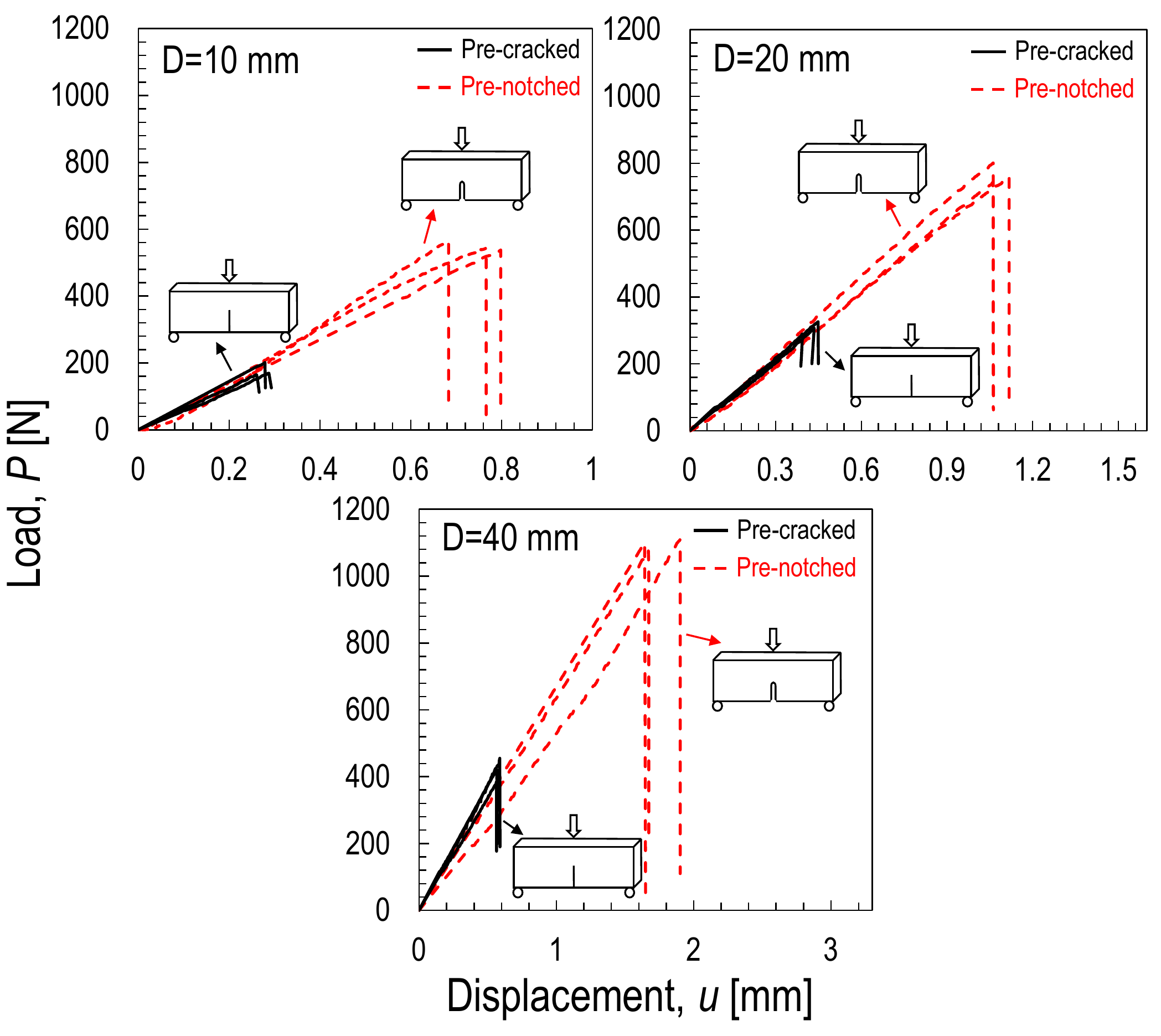}
\caption{Experimental load-displacement curves for pre-cracked and pre-notched specimens of different sizes.}
\label{fig:loaddis}
\end{figure}

The load-displacement curves of the three-point bending tests are plotted in Figure \ref{fig:loaddis}. As can be noted, for both pre-cracked and pre-notched specimens, the mechanical behavior is linear up to the peak load which is followed by unstable crack propagation. This is an indication of pronounced brittle behavior for all the investigated specimens. Further, as the figure shows, the stiffness of the specimens is not affected by the sharpness of the notch which is not surprising given the significantly large aspect ratios. In contrast, the peak load of the pre-notched specimens is approximately $3$ times higher that the one of the pre-cracked specimens for all the investigated sizes. This difference, which agrees with previous investigations \cite{xiao,cayard,alicia,souza,kuppu}, will be discussed in the following sections.

The crack or notch length, peak load, and nominal strength $\sigma_{Nc}=3P_cL/2tD^2$ for all the geometrically scaled specimens tested in this work are tabulated in Table \ref{tab:notchcrackloadstress}. In the definition of nominal strength, $P_c$ is the critical load, $t$ is the thickness of the specimens, $L$ is the span between the two supports, and $D$ is the width of the specimens.

\section{Analysis and Discussion}
\subsection{Estimation of the mode I fracture energy by LEFM}
\label{sec:LEFM}
In recent works \cite{Cory,Yao,Yao0,Yao1}, the effect of the Fracture Process Zone (FPZ) size on the fracturing behavior of thermoset nanocomposites was investigated leveraging size effect testing and analysis. It was shown that, Linear Elastic Fracture Mechanics (LEFM) provides a very accurate description of the fracture scaling in epoxy for typically laboratory-scaled, pre-cracked specimens as suggested by the ASTM D5045-99 \cite{ASTM_SENB}. This confirms that, for the pure epoxy and sufficiently large specimens, the FPZ size has a negligible effect. On the other hand, several investigations \cite{kinloch,lazzarin,lazzarin2,atzori,atzori2,Cusatis2} showed that, when the notch radius is sufficiently smaller than the Irwin's characteristic length, the notch has the same effects on the ultimate failure load as a crack of the same length. In this study, the notch radius b=$0.2$ mm is significantly smaller than the Irwin's characteristic length $l_{ch}=E^*G_{F}/f_{t}^2$ $\approx$ $0.75$ mm where $E=2263$ MPa, $\nu=0.35$ and $f_{t}=51.6$ MPa \cite{Cory}.



\begin{figure} [!ht]
\center
\includegraphics[scale=0.9]{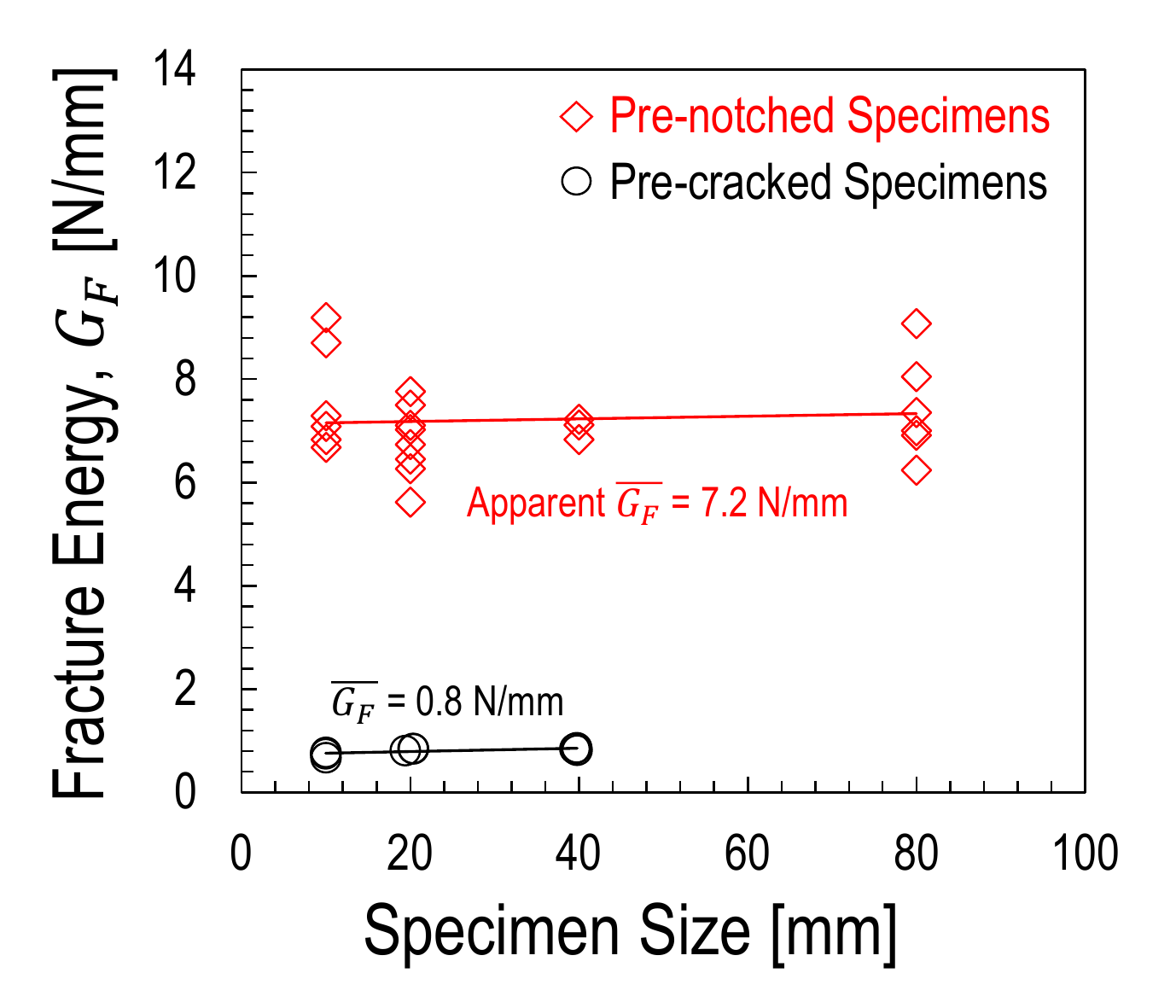}
\caption{Fracture energy estimated from LEFM for pre-cracked and pre-notched specimens of different sizes.}
\label{fig:LEFM}
\end{figure}

In light of the foregoing considerations, both the pre-cracked and pre-notched specimens can be analyzed by means of LEFM, leveraging the equations for a cracked specimen \cite{ASTM_SENB}:
\begin{equation}
G_{F}\left(\alpha_{0}\right)=\frac{\sigma_{Nc}^2D}{E^*}g(\alpha_{0})
\label{eq:GF}
\end{equation}
where $\alpha_{0}=a_{0}/D=$ normalized initial crack length, $\sigma_N=3PL/2tD^2=$ nominal stress, $E^*= E$ for plane stress and $E^*= E/\left(1-\nu^2\right)$ for plane strain, $\nu$ is the Poisson's ratio, and $g\left(\alpha_{0}\right)=$ dimensionless energy release rate which can be easily calculated following the procedure described in \cite{bazant1996_1,bazant1998_1,Cory,Yao,Yao0,Seung,Deleo,Cusatis,salviato2016_1}. 
The mode I fracture energy calculated by means of Eq.(\ref{eq:GF}) is shown in Figure \ref{fig:LEFM} for the different specimen sizes and notch types. Not surprisingly, the fracture energy calculated by LEFM is not affected by the specimen size. This is a confirmation that, for all the specimens investigated in this work, the nonlinear damage in the Fracture Process Zone (FPZ) did not affect the structural behavior significantly and linear theories such as LEFM can be used to provide a first estimate of the fracture properties.

On the other hand, the results showed in Figure \ref{fig:LEFM} indicate a very significant effect of the type of notch, the fracture energy of the pre-notched specimens being approximately $10$ times higher than the one of the pre-cracked specimens. This result, abundantly confirmed in the literature \cite{xiao,cayard,alicia,souza,kuppu}, cannot be explained by LEFM since, according to this theory, the notched specimens investigated in this work should fail at the same load as the cracked ones and the fracture energy should be a material property not affected by any geometrical feature.


In \cite{xiao,cayard,alicia,souza,kuppu}, the higher apparent fracture energy was ascribed to the emergence of residual, plastic stresses during the pre-notching process. However, a direct validation of this statement was never provided. The following section focuses on this particular aspect and shows that these hypothetical residual stresses should be unrealistically high to justify the difference in fracture energy reported in the present work and in the literature \cite{xiao,cayard,alicia,souza,kuppu}. 

\subsection{Residual stresses} 
\begin{figure} [!ht]
\center
\includegraphics[scale=0.4]{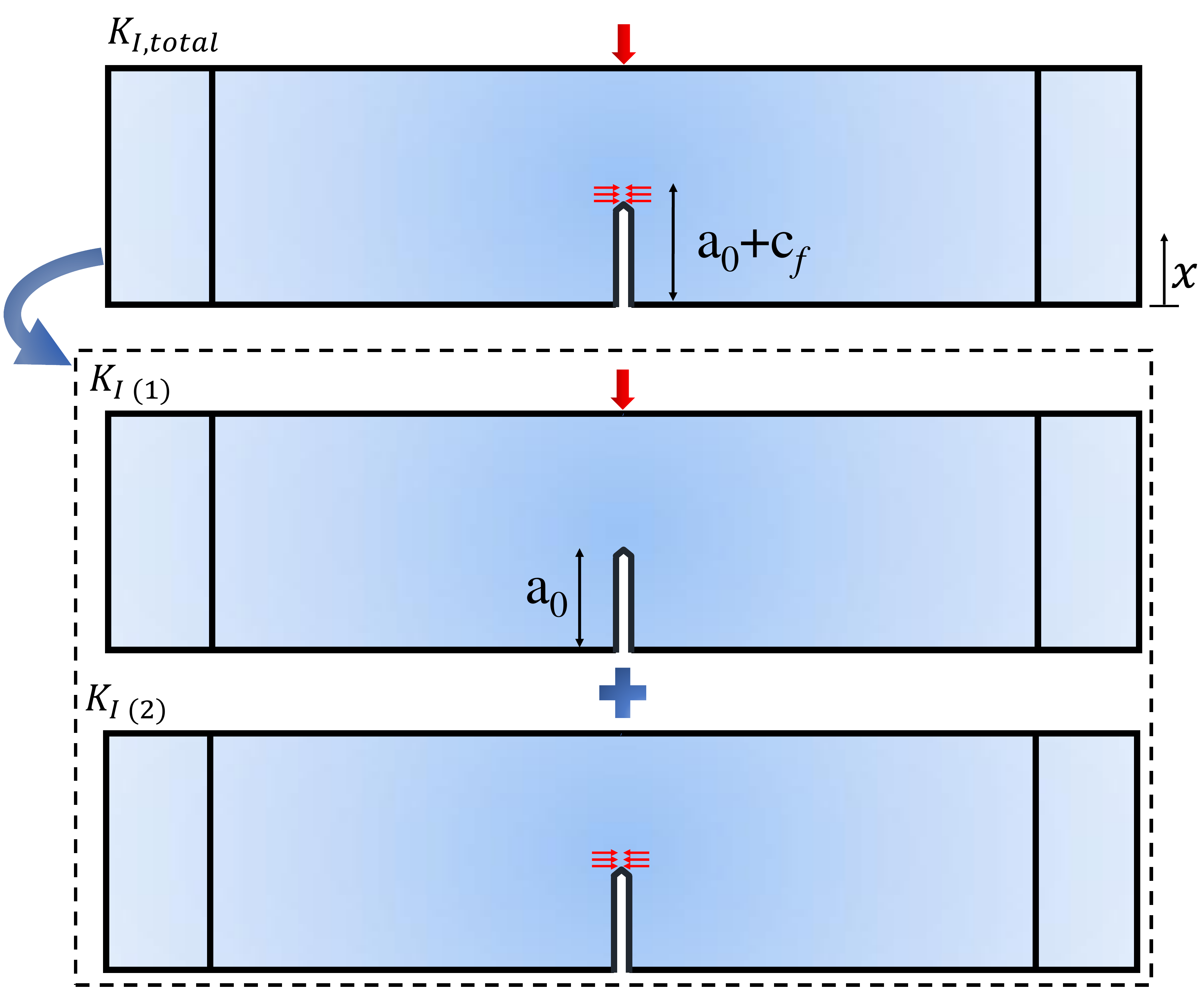}
\caption{Schematic description of the boundary conditions used for the estimation of the residual stresses by the superposition principle.}
\label{fig:residual}
\end{figure}

To check the possibility that the higher apparent fracture energy of pre-notched specimens is related to the presence of residual stresses as generally accepted in the literature, a simple analysis can be conducted within the framework of the Linear Elastic Fracture Mechanics (LEFM). Leveraging the superposition principle, the total mode I Stress Intensity Factor (SIF) can be calculated summing the effects of the applied load P and the residual stress distribution ahead of the crack tip:
\begin{equation}
K_{I(1)}+K_{I(2)}=K_{I,total}
\label{eq:residualone}
\end{equation}
where, as illustrated in Figure \ref{fig:residual}, $K_{I(1)}$ refers to the SIF associated to the concentrated load on the middle top of the specimen without the effect of residual stresses while $K_{I(2)}$  refers only to compressive stresses acting on the equivalent Fracture Process Zone of length $c_{f}$ \cite{Cory,Yao,Yao0,Seung,Deleo,Cusatis,bazant1996_1,bazant1998_1,salviato2016_1}. The length $c_{f}$ is proportional to the Irwin's characteristic length $l_{ch}$ and it is defined as an additional equivalent crack length required to capture the nonlinear effects of the FPZ. 

\begin{figure} [!ht]
\center
\includegraphics[scale=0.9]{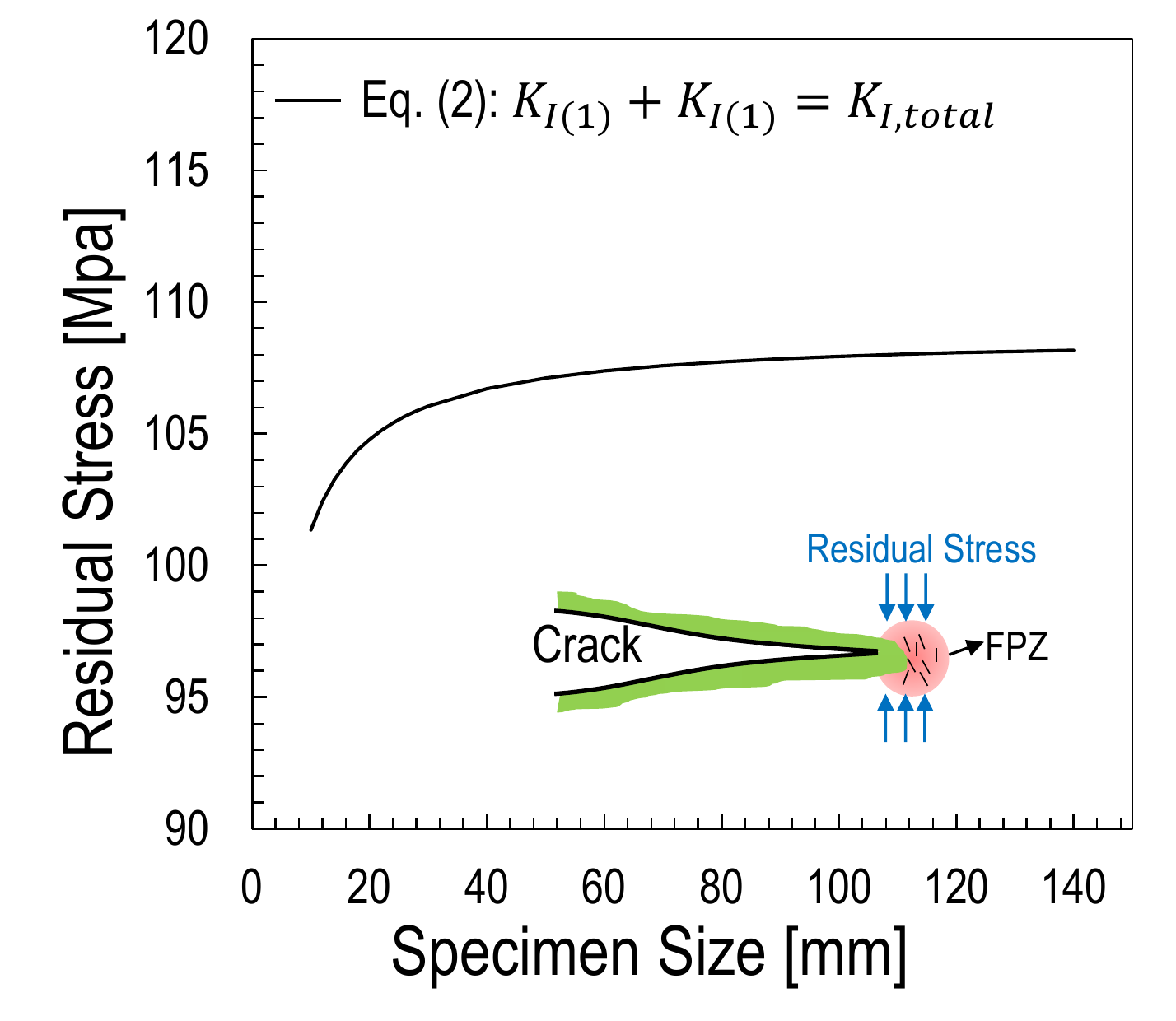}
\caption{The residual stress estimated from equivalent linear elastic fracture mechanics, Eq.(\ref{eq:residualone}).}
\label{fig:residualresult}
\end{figure}

The stress intensity factor $K_{I(1)}$ can be expressed as $\sqrt{E^*G}$ for a plane strain condition whereas $K_{I(2)}$ has the following expression \cite{Irwin}:
\begin{equation}
K_{I(2)}=-\int_{a_{0}}^{a_{0}+c_{f}}\frac{2\sigma_{\theta}(x)}{\sqrt{\pi a_{0}}}\frac{f(\frac{x}{a_{0}},\frac{a_{0}}{D})}{(1-\frac{a_{0}}{D})^{3/2}[1-(\frac{x}{a_{0}})^2]^{1/2}} \mbox{d}x
\label{eq:residualtwo}
\end{equation}
    where $\sigma_{\theta}(x)$ is the magnitude of the compressive stresses applied on the equivalent FPZ, $f(\frac{x}{a_{0}},\frac{a_{0}}{D})$ is the dimensionless geometry function \cite{Irwin}, and $x$ is the distance from the bottom surface of the SENB specimen as shown in Figure \ref{fig:residual}. At incipient crack onset, the stress intensity factor $K_{I,total}$ is ought to be equal to the fracture toughness of the pre-cracked specimens measured from the experiments. Accordingly, with the assumption that $a_{0}/D \approx 0.5$, $c_{f} \approx 0.5l_{ch}$ \cite{Cusatis,Deleo}, and the residual stresses are uniformly distributed, the magnitude can be estimated. As illustrated in Figure \ref{fig:residualresult}, the residual stress for the pre-notched specimens investigated in this work is higher than $100$ MPa which seems unrealistically high to be created by sawing during the pre-notching process.


To shed more light on the possible effects of residual stresses on the fracturing behavior, additional tests on pre-notched specimens were conducted. However, this time, a sharp razor blade was pre-inserted into the specimens during the manufacturing process and eventually removed after the curing of the epoxy resin to create the notch. Thanks to this procedure, notch tip radii similar to the ones of the sawed specimens were obtained without the possible emergence of plastic residual stresses.

\begin{figure} [!ht]
\center
\includegraphics[scale=0.7]{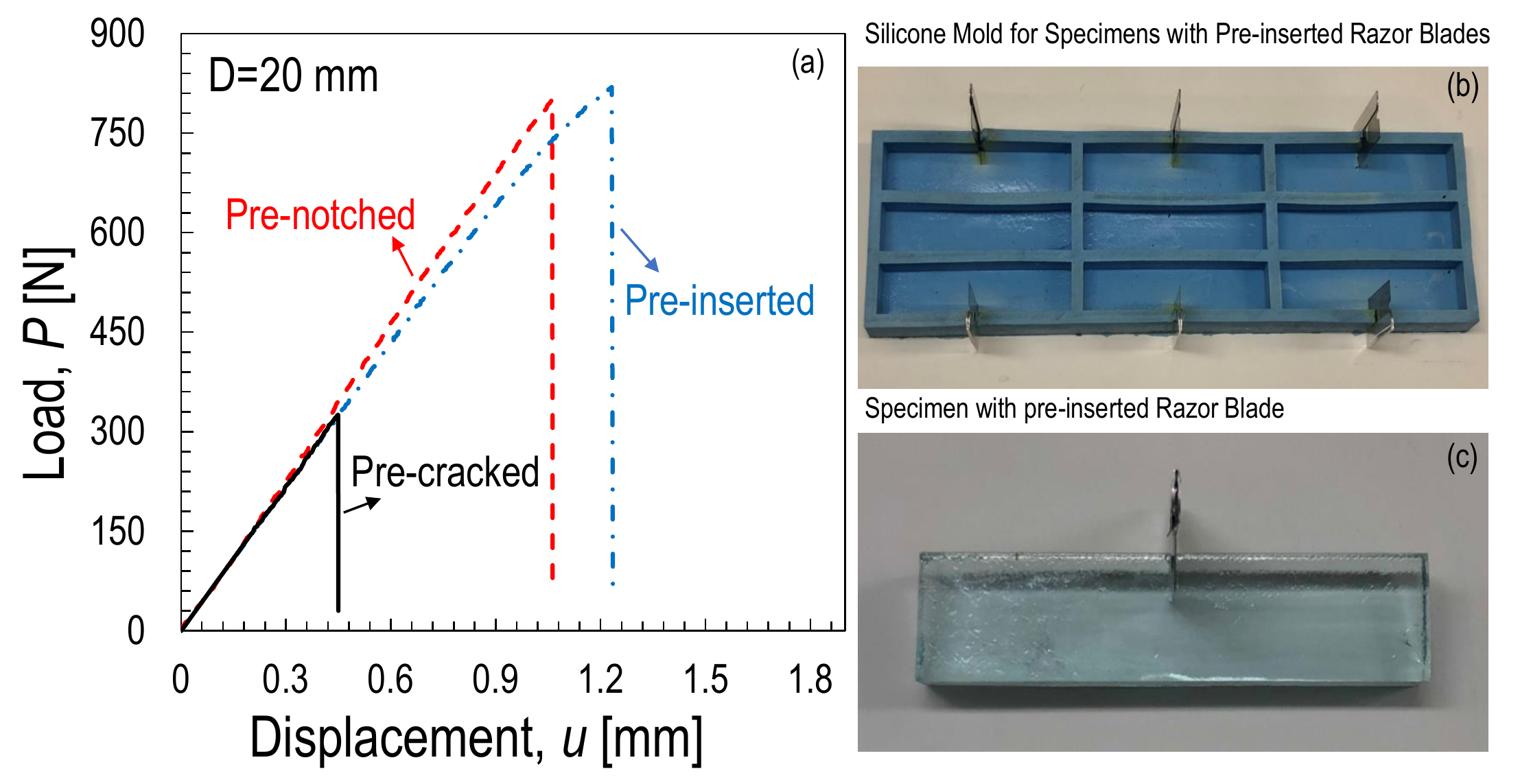}
\caption{(a) experimental load-displacement curves for pre-cracked and pre-notched specimens, and specimens with the notch made by a pre-inserted razor blade; (b) silicone molds for the manufacturing of the specimens with the pre-inserted razor blade; (c) typical specimens with the pre-inserted razor blade right after curing. Note that the blade was removed before the tests.}
\label{fig:preinsert}
\end{figure}

Figure \ref{fig:preinsert} compares the load displacement curves for specimens with pre-cracks made by tapping, with pre-notches made by sawing, and with pre-notches made by pre-inserting a razor blade. As can be noted, all the notched specimens exhibit a significantly larger peak load compared to cracked specimens. More importantly, the mechanical behavior of specimens with notches made by pre-inserting a razor blade is basically identical to the one reported for the case of sawed notches. 
Considering that the former cannot feature any residual stress, it is concluded that plastic residual stresses are not likely the dominant reason of the higher apparent fracture energy of pre-notched specimens. A possible explanation of the true cause is presented next leveraging cohesive fracture mechanics.

\subsection{Cohesive zone modeling of thermoset polymers}

In this section a two-scale cohesive zone model is proposed to capture the fracturing behavior of both pre-cracked and pre-notched specimens seamlessly. The main observations that led to the development of the model are described next along with the comparison with the experimental data.

\subsubsection{Bi-linear cohesive law for large crack opening displacements} 
\label{sec:Bilinear}
\begin{figure} [!ht]
\center
\includegraphics[scale=0.55]{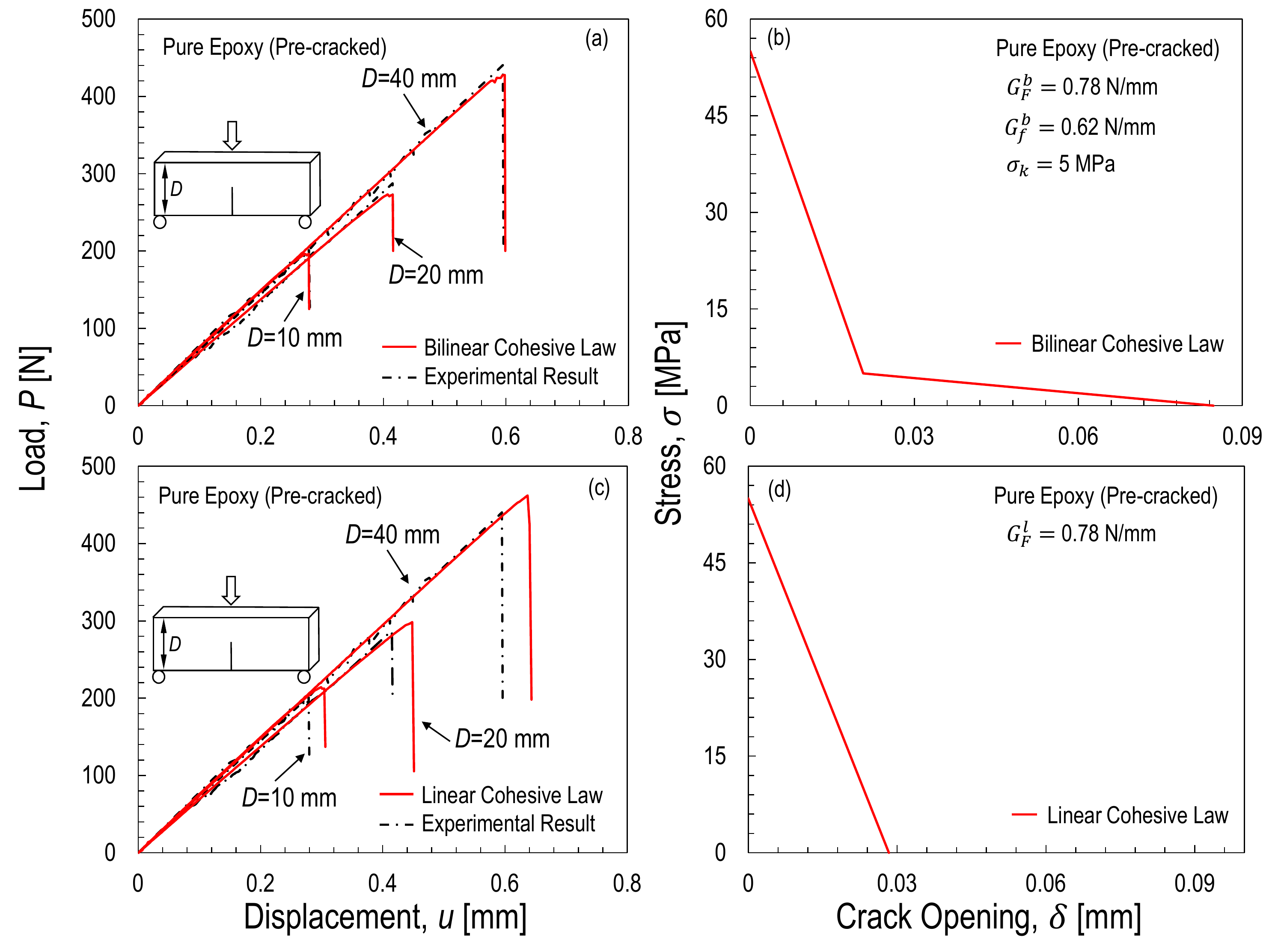}
\caption{Load-displacement curves vs. cohesive zone model featuring a linear and bi-linear cohesive law for geometrically-scaled, pre-cracked specimens of different sizes.}
\label{fig:bilinearresult}
\end{figure}

In previous investigations on pre-cracked, geometrically-scaled specimens made of thermoset resin \cite{Yao,Yao0}, it was found that the cohesive behavior of this material is best described by a bi-linear law. This is evident from Figures \ref{fig:bilinearresult}a-b showing that the bi-linear cohesive law enables the matching of the nominal strength $\sigma_{Nc}$ with errors always lower than $7$\% for all the specimen sizes investigated. In contrast, the errors are in the order of $30\%$ in case of a linear cohesive law (see Figures \ref{fig:bilinearresult}c-d).

As shown in Figure \ref{fig:bilineartrilinear}b, the bi-linear cohesive law can be described through four parameters: (a) tensile strength $f_t$, (b) initial fracture energy,  $G_f^{b}$, which represents the area under the initial segment of the bi-linear cohesive law; (c) total fracture energy, $G_F^b$, which is the total area under the bi-linear cohesive law; (d) change-of-slope stress, $\sigma_k$, which is the value of stress at the intersection of the initial and tail segment. By matching the size effect data on the structural strength of the pre-cracked specimens, the cohesive parameters of the bi-linear cohesive law are tabulated in Table \ref{tab:parameter}.

However, notwithstanding the excellent agreement with the data on pre-cracked specimens, it is impossible to match the experimental results on the pre-notched specimens by means of the foregoing cohesive law. A relatively good agreement is possible only by increasing the total fracture energy to $G_{F}^b=7.2$ N/mm, the value corresponding to the energy estimated by LEFM. However, this value would lead to a significant over-prediction of the structural strength of the pre-cracked specimens. Considering that the cohesive law should be a material property, the discrepancy between the results for pre-cracked and pre-notched specimens must be explained using the same cohesive law. A possible solution is presented next.


\subsubsection{Two-scale cohesive law} 
\label{sec:trilinear}
In a recent publication \cite{Cusatis2}, Di Luzio and Cusatis investigated the fracturing behavior of structures featuring blunt notches and characterized by materials exhibiting a linear cohesive law. In their numerical study they found that when the notch tip radius is approximately equal to the Irwin's characteristic length $l_{ch}=E^*G_f/f_t^2$ the failure occurs when the maximum stress at the notch approximately reaches the material strength, i.e. $\sigma_{Nc}=f_t/k$ with $k=$ stress concentration factor. On the other hand, for sufficiently sharp notches, the fracture is driven by the formation of a Fracture Process Zone (FPZ) in front of the notch which ultimately propagates and leads to the final failure.

\begin{figure} [!ht]
\center
\includegraphics[scale=0.75]{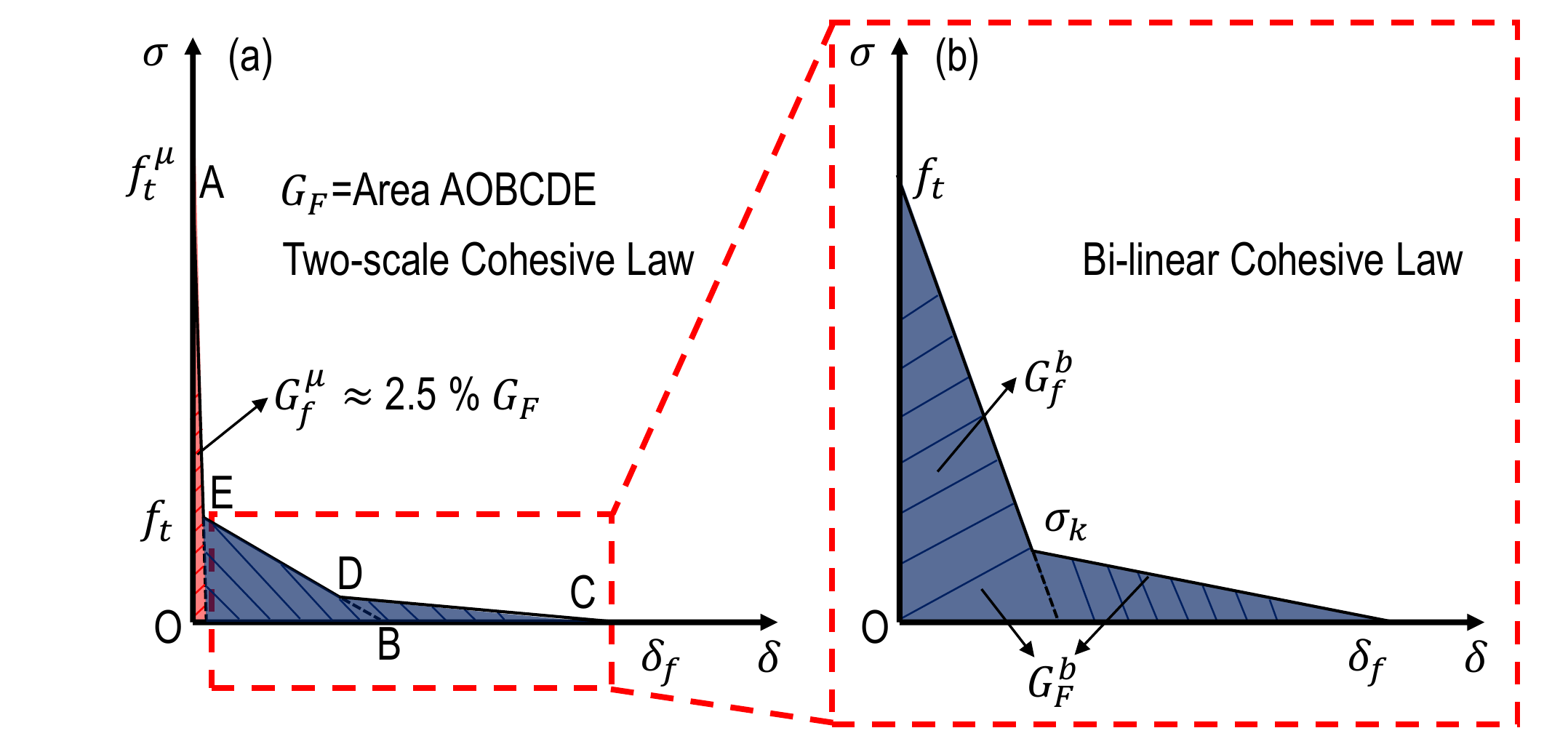}
\caption{(a) Schematic of the two-scale cohesive law proposed in this study describing the cohesive stresses associated to micro to macro crack opening displacements; (b) Bi-linear portion describing the fracturing for large crack opening displacements.}
\label{fig:bilineartrilinear}
\end{figure}

Based on the foregoing considerations, a cohesive zone model should be able to capture the behavior of notched and cracked specimens seamlessly provided that some particular features are added to the cohesive law. In this work, it is assumed that the cohesive behavior can be described by a two-scale cohesive law as represented in Figure \ref{fig:bilineartrilinear}a. As shown in the figure, $f_t^{\mu}$ represents the initial strength whereas $G_f^{\mu}$ is the initial fracture energy or, in other words, the area under the first linear branch of the curve. The rest of the curve is identical to the cohesive law identified by fracture tests on geometrically-scaled SENB specimens \cite{Yao,Yao1} and is characterized by $f_t$ = macroscopic strength, $\sigma_k=$ stress at the third change of slope, $G_f=$ initial macroscopic fracture energy (area AOBDE), and the total fracture energy $G_F$ (area AOBCDE).

The proposed cohesive law must differ from the linear one investigated in \cite{Cusatis2} to capture the peculiar cohesive behavior of thermoset polymers. Until the crack opening displacements are in the sub-micron regime, the resulting cohesive stresses are equivalent to the ones predicted by a linear cohesive law of strength $f_t^{\mu}$ and total energy $G_f^{\mu}$. For larger opening displacements, the cohesive law becomes equivalent to the bi-linear law that provided an excellent agreement with the experimental data on pre-cracked specimens. Since the first portion of the cohesive law describes the cohesive stresses at the microscale while the rest of the cohesive curve captures the behavior for larger displacements, the proposed cohesive law is characterized by two very distinct length scales. For this reason, in this and in future contributions the model will be referred to as a \textit{two-scale cohesive model}.

In this work, it is assumed that for sub-micron crack opening displacements the cohesive strength $f_t^{\mu}$ is about $300$ MPa and the cohesive stresses decrease linearly and very steeply with increasing crack openings. This captures the fact that, due to statistical size effect, the microscopic strength of the polymer can be several times higher than the one measured from macroscopic tests \cite{sanita,Chevalier}. At the same time, the steep initial part of the cohesive curve, leading to an initial fracture energy of only about $2.5\%$ of the total one, captures the lower energy dissipation occurring at the sub-micron scale. The value of about $300$ MPa is estimated from the failure loads of the pre-notched specimens. In fact, since the initial fracture energy $G_f^{\mu}$ is only a fraction of the total energy dissipated and the initial strength $f_t^{\mu}$ is significantly larger than in macroscopic tests, the Irwin's characteristic length $l_{ch}^{\mu}=E^*G_f^{\mu}/\left(f_t^{\mu}\right)^2$ related to the initial formation of the FPZ is significantly smaller than the width of the notches investigated in this work. Accordingly, following \cite{Cusatis} the fracturing behavior of the notched specimens must depart from the one of the cracked specimens and the nominal stress at failure is determined by the elastic limit condition $\sigma_{Nc}=f_t^{\mu}/k$. It is worth mentioning here that, thanks to the foregoing considerations, the initial strength can be determined very precisely. Small variations on the value proposed in this work would lead to significant changes on the predicted structural strength.


Having clarified the reason for the very high strength at the microscale, another question needs to be answered: why the cohesive stresses of the initial portion of the cohesive curve must decrease so steeply? The reason is that, if the initial slope of the cohesive curve was milder, it would not be possible to capture the size effect data on pre-cracked specimens. In the presence of a crack or a sufficiently sharp notch, the stress intensity at the tip at incipient failure would always lead to elastic stresses that are greater then the microscale strength. Accordingly, since the cohesive stresses decrease very quickly in the initial part of the cohesive curve, a cohesive FPZ can emerge and develop up to the second portion of the tri-linear cohesive curve. The cohesive law becomes equivalent to the bi-linear law of strength $f_t$ and total fracture energy $G_F^b$ since the initial energy $G_f^{\mu}$ is only about $2.5\%$ of $G_F$ or, in other words, $G_F\approx G_F^b\approx 0.80$ N/mm. Thanks to the foregoing considerations it is possible to estimate the initial slope of the two-scale cohesive model very precisely by testing notched specimens of various notch tip radii and matching the experimental data by means of the two-scale cohesive model. Such a comprehensive experimental campaign is beyond the scope of the present work and will be the subject of future contributions. In this study, the initial fracture energy was calibrated against one notch tip radius only, leading to $G_f^{\mu}=$0.02 N/mm.

Finally, one last question remains: why the foregoing cohesive behavior was not found before from tests on pre-cracked specimens? The answer lies in the particular shape of the cohesive curve. Since the initial fracture energy is only a negligible portion of the total fracture energy and, in laboratory-scale cracked specimens the cohesive stresses always overcome the initial portion of the cohesive law, it is not possible to characterize $f_t^{\mu}$ and $G_f^{\mu}$ unless micro-metric specimens are tested. In fact, any tests on large specimens would provide an estimate of $G_F$ which takes approximately the same value as $G_F^b$. It is worth mentioning here that micro-tests are very challenging and are generally affected by significant uncertainties. However, based on the results of this work, a valid alternative is to test both cracked and pre-notched specimens and use the results on the latter configuration to characterize $f_t^{\mu}$.

\begin{figure} [!ht]
\center
\includegraphics[scale=0.6]{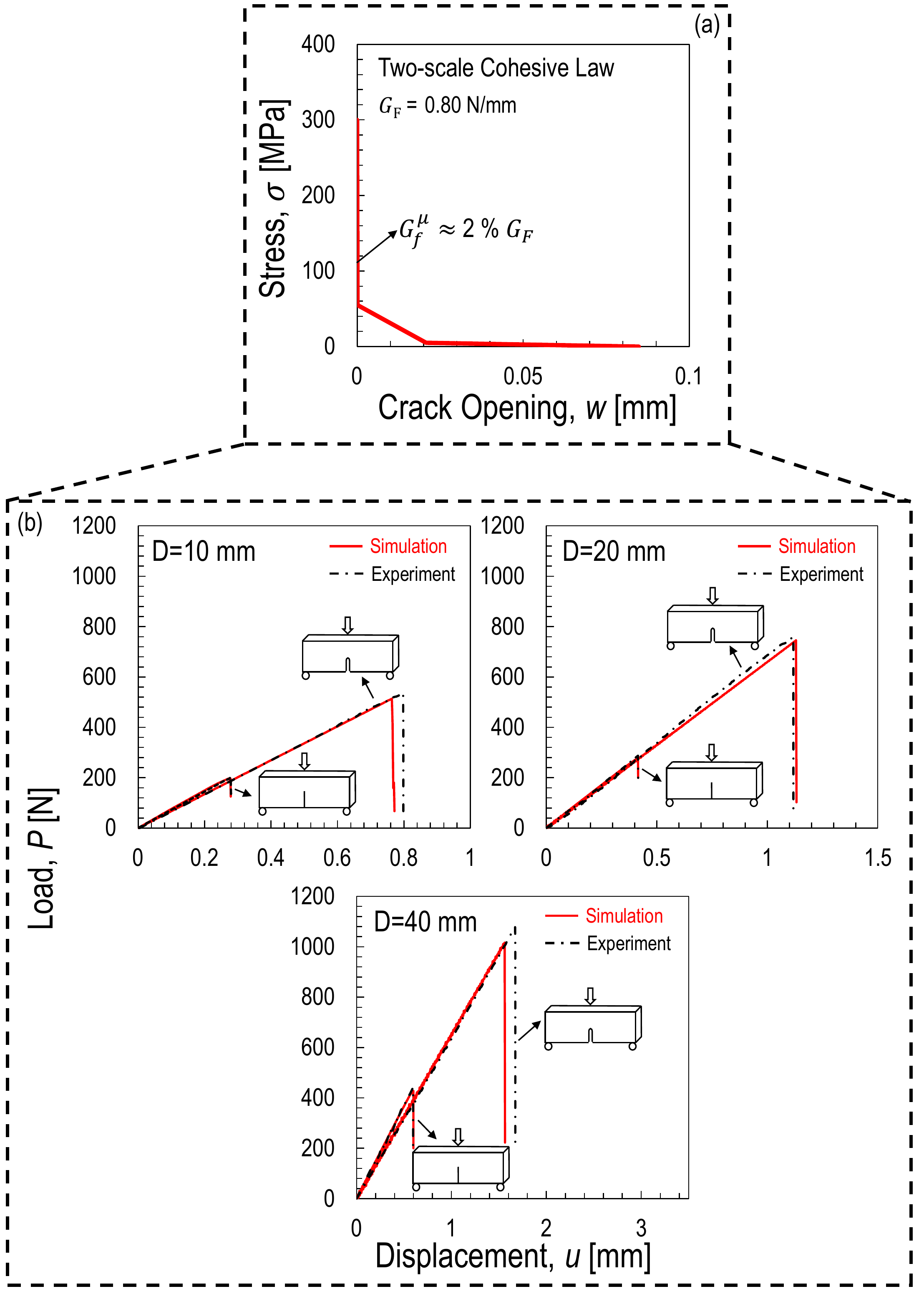}
\caption{Cohesive zone modeling results: (a) calibrated cohesive law; (b) load-displacement curves vs. two-scale Cohesive Zone Model (CZM) simulations for both pre-cracked and pre-notched specimens of different sizes.}
\label{fig:trilinearresult}
\end{figure}

To verify the foregoing assumptions, both pre-cracked and pre-notched specimens were simulated in ABAQUS/Explicit 2017. The models combined 4-node bi-linear plain strain quadrilateral elements (CPE4R) with a linear elastic isotropic behavior and 4-node two-dimensional cohesive elements (COH2D4) with the traction-separation law shown in Figure \ref{fig:trilinearresult} to model the crack. The parameters of the two-scale cohesive model that provided the best matching with the experimental data are given in Table \ref{tab:parameter}.


Figure \ref{fig:trilinearresult} shows a comparison between the experimental load-displacement curves and simulations by the two-scale cohesive model. As can be noted the model successfully matches the experimental curves of both the pre-notched and pre-cracked specimens of different sizes. The most remarkable aspect related to these results is that the excellent matching was possible leveraging the \textit{same cohesive law} which can now be treated as a material property. 

\begin{figure} [!ht]
\center
\includegraphics[scale=0.5]{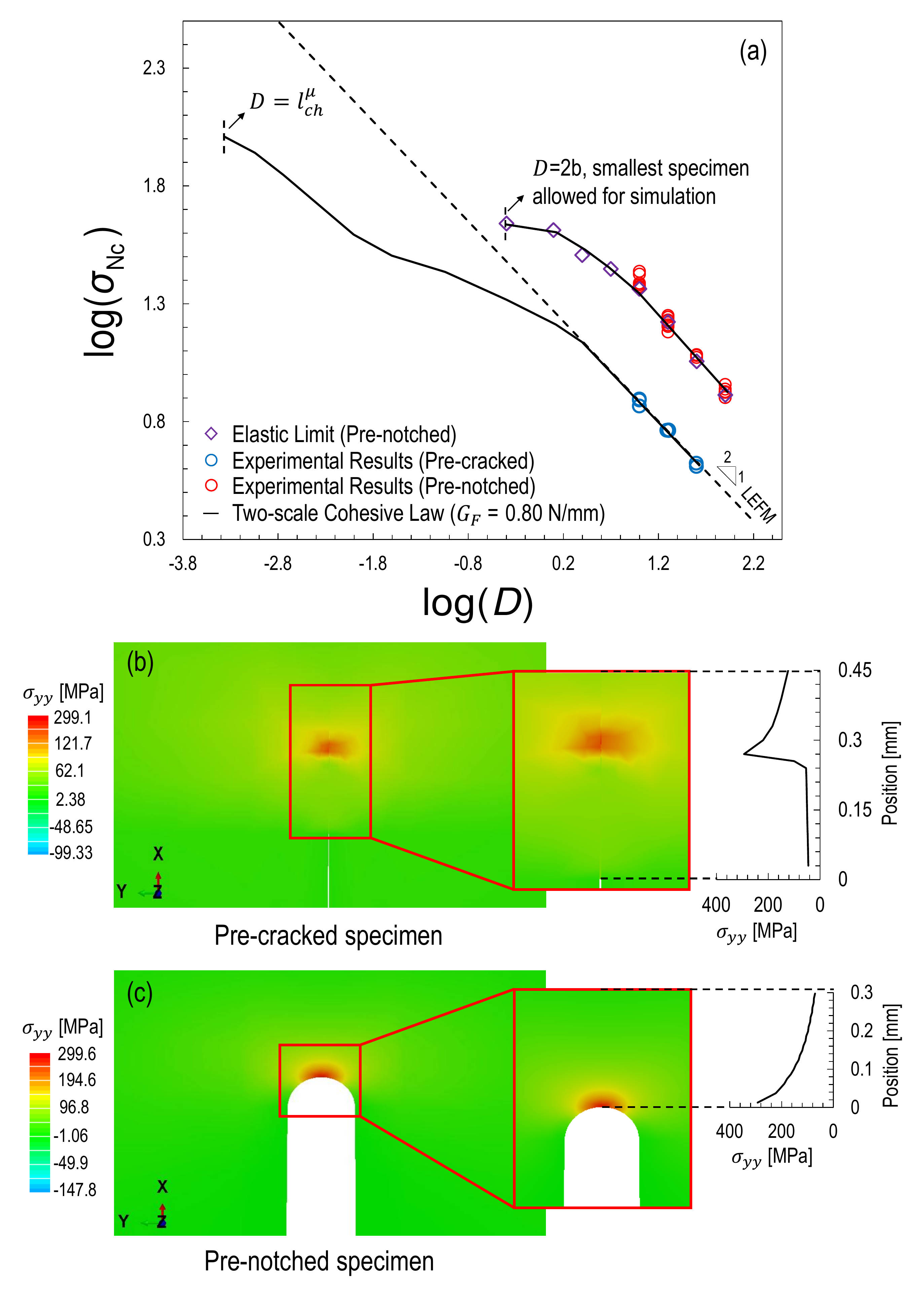}
\caption{Experimental results vs. two-scale Cohesive Zone Model (CZM) simulations on the structural strength of pre-cracked and pre-notched specimens of different sizes.}
\label{fig:structuralstrength}
\end{figure}

It is interesting to plot the structural strength $\sigma_{Nc}$ as a function of the structure size $D$ in double-logarithmic scale. As can be noted from Figure \ref{fig:structuralstrength}, the computational model is in excellent agreement with the experimental data for all the sizes and types of notch. Further, the model seems to capture very well the size effect in both cracked and notched specimens. 

In case of cracked specimens the radius at the crack tip is extremely small, approximately zero. For sufficiently large specimens, the FPZ size is negligible compared to the structure size and, as predicted by LEFM, the structural strength scales with $D^{-1/2}$. For decreasing sizes, the fraction of the structure occupied by the nonlinear FPZ becomes larger and larger thus affecting the fracturing behavior. The structural strength departs from the values predicted by the LEFM which is a linear theory and, inherently, ignores the cohesive stresses in the FPZ. For sufficiently small geometrically-scaled structures the nominal strength tends to the plastic limit.

On the other hand, the two-scale model predicts a significant size effect also for the notched specimens as shown by the predictions plotted in Figure \ref{fig:structuralstrength}. For the cases shown in the Figure, all the geometrical features of the structure are geometrically-scaled except for the thickness and notch tip radius, $b$. Accordingly, the aspect ratio of the notch $a_0/b$ increases along with the structure size, $D$. As can be noted, the structural strength decreases significantly with $D$, this size dependence becoming less and less significant for decreasing structure sizes. This is shown in the figure by the decreasing slope of the curve of the structural strength for lower $D$ values. The diamond symbols in Figure \ref{fig:structuralstrength} represent the elastic limit for the various structure sizes, $\psi\left(b,D\right)=f_t^{\mu}/k\left(b, D\right)$. In this expression, the stress concentration factor is a function of $D$ since the notch tip radius is not geometrically scaled. As can be noted, the function $\psi\left(b,D\right)$ agrees very well with the experimental data. This confirms that, for the notched specimens investigated in this work, all the failures happened by approximately reaching the elastic limit of the structure.



 Finally, it is interesting to compare the cohesive stresses at the condition of incipient failure for the case of the specimens weakened by cracks and notches of equal depth considered in this work. As shown in Figures \ref{fig:structuralstrength}b,c, the cohesive stresses in these two cases differ significantly, in accordance with the observations that motivated the proposed two-scale cohesive model. For notches, the FPZ is developed only partially, with the minimum cohesive stress being significantly larger than $f_t$. This means that, in agreement with the theory, only the initial part of the two-scale cohesive model governs the behavior of the notched structures investigated in this work. In contrast, in case of pre-cracked specimens, the FPZ is far more developed and the minimum cohesive stress at the tip is generally much lower than $f_t$. This means that the second part of the cohesive law is entered and the cohesive behavior is equivalent to a bi-linear cohesive law of total fracture energy $G_F^b \approx G_F$. All the foregoing observations confirm firmly the validity of the two-scale cohesive model.

 Furthermore, based on the two-scale model proposed in this work, some considerations on the applicability of LEFM for a course estimate of the failure load can be made. In case of pre-cracked specimens, LEFM can only be applied when the FPZ size is negligible compared to the structure size. If this condition is met for micro-metric structures, then the fracture energy to be used in the calculations is $G_f^{\mu}$. This means that to predict e.g. the onset of a microcrack in a polymer matrix composite by LEFM, one has to first verify that the micro Irwin's characteristic length $l_{ch}^{\mu}$ is significantly smaller than the smallest geometrical features (e.g. fiber diameter, inter-fiber distance etc). Then, if the condition is met, LEFM can used provided that the microscale fracture energy $G_f^{\mu}$ is used in the calculations. On the other hand, if the condition of negligible FPZ size is met for structures of size $D \gg l_{ch}=E^*G_F/f_t$, the total fracture energy $G_F$ should be used in the calculations.   
 
 In case of pre-notched specimens, LEFM can only be applied when the notch tip radius is extremely small, smaller than the micro Irwin's characteristic length. If this condition is met, then the same observations on the applicability of LEFM to pre-cracked structures hold.


\section{Conclusions}
This study investigated the fracturing behavior of thermoset polymer structures featuring cracks and sharp u-notches. Based on the results obtained in this work, the following conclusions can be elaborated:  

1. the fracturing behavior of thermoset polymer structures featuring cracks is distinctively different than the one of structures weakened by u-notches. Although this result is not surprising for blunt notches with large tip radii, it was not expected for the notches investigated in this work which featured a tip radius significantly smaller than the Irwin's characteristic length of the material;

2. for the cases studied in this work, the direct use of LEFM to estimate the mode-I fracture energy of the material provides values for the notched specimens that are about $10$ times larger than the ones estimated from cracked specimens ($7.2 $ N/mm for notched specimens vs $0.78$ N/mm for cracked specimens); 

3. in contrast to what has been proposed in the literature, plastic residual stresses induced by the manufacturing of the notch cannot be the dominant cause of such a high apparent fracture toughness. In fact, the residual stresses that would be required to justify such a high toughness are larger than $100$ MPa and it seems unlikely that they can be generated by the sawing procedure. This statement is supported by additional tests on specimens with notches created by pre-inserting a sharp razor blade. Even if in such case plastic stresses cannot be induced, the failure loads were similar to those of the other notched specimens;

4. to capture the behavior of both the cracked and notched specimens, a \textit{two-scale cohesive model} is proposed. For sub-micron crack opening displacements, the cohesive strength is about $300$ MPa and the cohesive stresses decrease linearly and very steeply with increasing crack openings. When the cohesive stresses reach the strength estimated from typical laboratory-scale specimens (about $55$ MPa) and the opening displacements are beyond the sub-micron range, the slope of the cohesive law becomes milder. Finally, the cohesive law ends with another linear branch starting when the stresses reach $5$ MPa. This last part of the cohesive law is characterized by a very slow decay of the cohesive stresses;


5. the proposed two-scale cohesive model captures the fact that, due to energetic-statistical size effect \cite{Le1,Le2,Le3,Salviato3,Salviato4}, the microscopic strength of the polymer can be several times higher than the one measured from macroscopic tests \cite{sanita,Chevalier}. At the same time, the steep initial part of the cohesive curve, leading to an initial fracture energy of only about $2.5\%$ of the total one, captures the lower energy dissipation occurring at the sub-micron scale; 

6. thanks to the foregoing model the different behavior of polymer structures featuring cracks and sharp u-notches can be easily explained. Since the initial fracture energy $G_f^{\mu}$ is only a fraction of the total energy dissipated and the initial strength $f_t^{\mu}$ is significantly larger than in macroscopic tests, the Irwin's characteristic length $l_{ch}^{\mu}=E^*G_f^{\mu}/\left(f_t^{\mu}\right)^2=5.4$ $\mu m$ related to the initial formation of the FPZ is significantly smaller than the width of the notches investigated in this work. Accordingly, the fracturing behavior of the notched specimens must depart from the one of cracked specimens and the nominal stress at failure is determined by the elastic limit condition \cite{Cusatis2}, i.e. $\sigma_{Nc}=f_t^{\mu}/k$ with $k=$ stress concentration factor. On the other hand, in laboratory-scaled cracked specimens the cohesive stresses in the FPZ at incipient failure always reach the second portion of the cohesive curve. In such a case, the Irwin's characteristic length $l_{ch}$ describing the FPZ is significantly larger since it depends on the total fracture energy $G_F\approx 40G_f^{\mu}$ and a significantly lower strength $f_t\approx f_t^{\mu}/6$. Accordingly, the failure behavior of macroscopic cracked specimens depends only on the total fracture energy and macroscopic strength; 

7. since the initial fracture energy is only a negligible portion of the total fracture energy and, in laboratory-scale cracked specimens the cohesive stresses always overcome the initial portion of the cohesive law, it is not possible to characterize $f_t^{\mu}$ and $G_f^{\mu}$ unless micro-metric specimens are tested. These tests are very challenging and are generally affected by significant uncertainties. Based on the results of this work, a valid alternative is to test both cracked and pre-notched specimens and use the results on the latter configuration to characterize $f_t^{\mu}$; 

8. the foregoing considerations are supported by the excellent agreement between the model predictions and experimental data on geometrically-scaled Single Edge Notch Bending (SENB) specimens. As can be noted from Figures \ref{fig:trilinearresult} and \ref{fig:structuralstrength}, the two-scale cohesive model is able to capture not only the load-displacement curves but also the related size effect on structural strength for both cracked and pre-notched specimens;

9. the foregoing results are of utmost importance for the design of microelectronic devices or polymer matrix composites. In fact, in these systems, the main damage mechanisms belong to the sub-micron scale and the fracturing behavior is dominated by the initial portion of the cohesive law since the opening displacements are not large enough to reach the second portion of the curve. The two-scale model proposed in this work and the related testing protocol, represent a first step towards the accurate yet simple simulation of thermoset polymers at the microscale.




\section*{Acknowledgments}
Marco Salviato acknowledges the financial support from the Haythornthwaite Foundation through the ASME Haythornthwaite Young Investigator Award. This work was also partially supported by the Joint Center For Aerospace Technology Innovation through the grant titled ``Design and Development of Non-Conventional, Damage Tolerant, and Recyclable Structures Based on Discontinuous Fiber Composites".


\newpage
\section*{Figures and Tables}

\newpage
\begin{table}[!ht]
\centering
\scalebox{0.7}{
\begin{tabular}{ccccc}
\hline
Specimen type &Specimen width (mm) &Crack or notch length (mm) &Max load (N) &Nominal strength (MPa) \\ \hline
Pre-notched &D=10 &5.00 &647.01 &28.12\\
Pre-notched &D=10 &5.00 &543.07 &23.60\\
Pre-notched &D=10 &5.00 &613.01 &26.65\\
Pre-notched &D=10 &5.00 &654.10 &28.40\\
Pre-notched &D=10 &5.00 &561.12 &24.39\\
Pre-notched &D=10 &5.00 &537.22 &23.34\\
Pre-notched &D=10 &5.00 &553.40 &24.03\\
Pre-notched &D=10 &5.00 &630.18 &23.37\\ 
Pre-notched &D=20 &10.00 &800.66 &17.49\\
Pre-notched  &D=20 &10.00 &758.70 &16.57\\
Pre-notched  &D=20 &10.00 &742.83 &16.22\\
Pre-notched  &D=20 &10.00 &779.50 &17.02\\
Pre-notched  &D=20 &10.00 &693.15 &15.14\\
Pre-notched  &D=20 &10.00 &731.95 &15.99\\
Pre-notched  &D=20 &10.00 &814.50 &17.79\\
Pre-notched  &D=20 &10.00 &774.82 &16.92\\ 
Pre-notched  &D=40 &20.00 &1100.05 &12.04\\
Pre-notched  &D=40 &20.00 &1077.89 &11.80\\
Pre-notched  &D=40 &20.00 &1108.72 &12.14\\ 
Pre-notched  &D=80 &40.00 &1580.11 &8.66\\
Pre-notched &D=80 &9.86 &1542.20 &8.45\\
Pre-notched &D=80 &15.86 &1653.11 &9.06\\
Pre-notched &D=80 &16.93 &1532.03 &8.40\\
Pre-notched &D=80 &17.68 &1455.30 &7.97\\ 
Pre-notched &D=80 &5.69 &1755.01 &9.62\\ \hline
Pre-cracked &D=10 &5.03 &169.33 &7.73\\
Pre-cracked &D=10 &4.96 &164.62 &7.47\\
Pre-cracked &D=10 &4.38 &201.01 &9.48\\ 
Pre-cracked &D=20 &7.63 &289.27 &7.91\\
Pre-cracked &D=20 &9.26 &306.58 &6.65\\
Pre-cracked &D=20 &9.28 &325.62 &6.61\\ 
Pre-cracked &D=40 &16.46 &455.31 &5.37\\
Pre-cracked &D=40 &17.27 &385.44 &4.93\\
Pre-cracked &D=40 &16.46 &455.31 &5.37\\ \hline
\end{tabular}}
\caption{Max load and nominal strength of pre-notched and pre-cracked specimens at different sizes.}
\label{tab:notchcrackloadstress}
\end{table}

\newpage
\begin{table}[!ht]
\centering
\scalebox{0.75}{
\begin{tabular}{ccccccccc}
\hline
Cohesive Law &$G_F$ (N/mm) &$G_F^{b}$ (N/mm) &$G_F^{l}$ (N/mm) &$G_f$ (N/mm) &$G_f^{b}$ (N/mm) &$f_t$ (MPa) &$f_t^{\mu}$ (MPa) &$\sigma_k$ (MPa) \\ \hline
Two-scale &0.8 &/ &/ &0.645 &/ &55 &300 &5\\
Bi-linear &/ &0.78 &/ &/ &0.625 &55 &/ &5\\
Linear &/ &/ &0.78 &/ &/ &55 &/ &/ \\ \hline
\end{tabular}}
\caption{Calibrated cohesive parameters for linear, bi-linear and two-scale cohesive laws.}
\label{tab:parameter}
\end{table}


\end{document}